\documentclass[12pt]{article}
\usepackage{amsmath,epsf,amssymb,cite}
\newtheorem{theorem}{Theorem}

\newtheorem{mainassertion}{Main Assertion}
\newtheorem{lemma}{Lemma}

\newif\iffigs\figstrue

\DeclareFontFamily{U}{rsf}{}
\DeclareFontShape{U}{rsf}{m}{n}{
  <5> <6> rsfs5 <7> <8> <9> rsfs7 <10-> rsfs10}{}
\DeclareMathAlphabet\Scr{U}{rsf}{m}{n}

\def\pplogo{\vbox{%\kern-\headheight\kern -29pt
\halign{##&##\hfil\cr&{%\sc
\ppnumber}\cr\rule{0pt}{2.5ex}&\ppdate\cr}
}}
\makeatletter
\def\ps@firstpage{\ps@empty \def\@oddhead{\hss\pplogo}%
  \let\@evenhead\@oddhead % in case an article starts on a left-hand page
}
\def\maketitle{\par
 \begingroup
 \def\thefootnote{\fnsymbol{footnote}}
 \def\@makefnmark{\hbox{$^{\@thefnmark}$\hss}}
 \if@twocolumn
 \twocolumn[\@maketitle]
 \else \newpage
 \global\@topnum\z@ \@maketitle \fi\thispagestyle{firstpage}\@thanks
 \endgroup
 \setcounter{footnote}{0}
 \let\maketitle\relax
 \let\@maketitle\relax
 \gdef\@thanks{}\gdef\@author{}\gdef\@title{}\let\thanks\relax}
\makeatother

\def\O{{\mathcal O}}
\def\C{{\mathbb C}}
\def\P{{\mathbb P}}

\def\Z{{\mathbb Z}}

\def\Hom{\operatorname{Hom}}

\def\tr{\operatorname{tr}}

\def\SO{\operatorname{SO}}

\def\SU{\operatorname{SU}}

\def\Sp{\operatorname{Sp}}
\def\Ad{\operatorname{Ad}}

\def\so{\operatorname{\mathfrak{so}}}
\def\su{\operatorname{\mathfrak{su}}}
\def\gu{\operatorname{\mathfrak{u}}}
\def\sp{\operatorname{\mathfrak{sp}}}

\def\CY{Calabi--Yau}

\def\cM{{\Scr M}}

\def\cD{{\Scr D}}

\def\cG{{\Scr G}}

\def\cMc{{\hfuzz=100cm\hbox to 0pt{$\;\overline{\phantom{X}}$}\cM}}
\def\barcD{{\hfuzz=100cm\hbox to 0pt{$\;\overline{\phantom{X}}$}\cD}}
\def\ff#1#2{{\textstyle\frac{#1}{#2}}}

\def\HS#1{{\mathbb{F}}_{#1}}
\def\Ist#1{{\mathrm{I}\vphantom{\mathrm{II}}}^*_{#1}}
\def\mf#1{\mathfrak{#1}}

\begin{document}
\setcounter{page}0
\def\ppnumber{\vbox{\baselineskip14pt
\hbox{DUKE-CGTP-00-02}
\hbox{OSU-Math-2000-3}
\hbox{hep-th/0002012}}}
\def\ppdate{February 2000} \date{}

\title{\LARGE Lie Groups, Calabi--Yau Threefolds,\\
   and F-Theory\\[10mm]}
\author{
Paul S. Aspinwall$^1$, Sheldon Katz$^2$, and David R. Morrison$^1$\\[10mm]
\normalsize $^1$Center for Geometry and Theoretical Physics, \\
\normalsize Box 90318, \\
\normalsize Duke University, \\
\normalsize Durham, NC 27708-0318\\[7mm]
\normalsize $^2$Department of Mathematics,\\
\normalsize Oklahoma State University, \\
\normalsize Stillwater, OK 74078\\[10mm]
}

{\hfuzz=10cm\maketitle}

\def\Large{\large}
\def\LARGE{\large\bf}

%\vskip 1cm

\begin{abstract}
The F-theory vacuum constructed from an elliptic Calabi--Yau threefold with
section yields an effective six-dimensional theory.  The Lie algebra of the
gauge sector of this theory and its representation on the space of massless
hypermultiplets are shown to be determined by the intersection theory of
the homology of the Calabi--Yau threefold.  (Similar statements hold for
M-theory and the type IIA string compactified on the threefold, where there
is also a dependence on the expectation values of the Ramond--Ramond
fields.)  We describe general rules for computing the hypermultiplet
spectrum of any F-theory vacuum, including vacua with non-simply-laced
gauge groups.  The case of monodromy acting on a curve of
$A_{\mathrm{even}}$ singularities is shown to be particularly interesting
and leads to some unexpected rules for how $2$-branes are allowed to wrap
certain $2$-cycles.  We also review the peculiar numerical predictions for
the geometry of elliptic Calabi--Yau threefolds with section which arise
from anomaly cancellation in six dimensions.
\end{abstract}

\vfil\break

%%%%%%%%%%%%%%%%%%%%%%%%%%%%%%%%%%%%%%%%%%%%%%%%%%%%%%%%%%%%%%%%

\section{Introduction}    \label{s:int}

The F-theory vacuum \cite{Vafa:F} constructed from  an elliptically
fibered \CY\ 
threefold $X$ with section determines an effective
theory with $N=(1,0)$ supersymmetry in six 
dimensions.
Such supersymmetric 
theories will have fields in hypermultiplets, vector supermultiplets
and tensor supermultiplets. (See, for example, \cite{SW:6d} for a
discussion of such theories.)

For any particular F-theory vacuum, the taxonomy of the
supermultiplets may be derived from the geometry 
of $X$ as an elliptic fibration via seemingly straightforward methods
in the case of the vector and tensor multiplets \cite{MV:F,MV:F2}.
The classification of the hypermultiplet content has always been a
little harder to carry out. Many methods have been proposed which allow
the hypermultiplets to be determined from the geometry in certain
cases 
\cite{KMP:enhg,W:MF,AG:sp32,BKV:enhg,Sad:anom,KV:hyp,IMS:5deg,CPR:mtor,DE:5d}.

The purpose of this paper is to outline a systematic approach to the
problem of 
determining the gauge symmetry and hypermultiplet content of a given
six-dimensional theory obtained from F-theory. (Note that as far as the
moduli space of hypermultiplets in concerned, our methods utilize the
associated type IIA compactification and thus
also apply directly to the compactification of M-theory on $X$ giving
an $N=1$ theory in five dimensions and to the compactification of the
type IIA string on $X$ to yield an $N=2$ theory in four dimensions,
provided that the expectation values of certain Ramond--Ramond fields have
been tuned appropriately.)

The methods we employ will not be particularly new but we will see
that the process of analyzing the gauge group and matter content can
be quite a bit more subtle than had previously been appreciated. 
In particular, the case of monodromy of the fibration leading to
non-simply-laced Lie algebras requires some care. A particularly awkward
case which has caused some confusion is when a $\Z_2$ monodromy acts on
a curve of $A_{\mathrm{even}}$ singularities, i.e., a curve of
$\mathrm{I}_{\mathrm{odd}}$ fibers in F-theory language. In this paper
we resolve this problem in agreement with an observation by
Intriligator and Rajesh in \cite{In:po} concerning anomaly cancellation.

In section \ref{s:lcy} we will show how many features of a Lie algebra
structure arise 
naturally from an elliptically fibered \CY\ threefold. This will allow
us to elucidate the method for determining the gauge algebra.
In section \ref{s:hyper} we discuss exactly how to analyze the
hypermultiplet content in the cases where the associated curves and
surfaces within the \CY\ threefold are smooth. We discuss the cases
where these curves and surfaces are singular in section
\ref{s:odd}. 
This section includes some unexpected rules we are forced to
adopt for $2$-brane wrapping. Although the results of this section are
less rigorous than the preceding section, we are able to give precise
results in many instances which can be extended to the general case under
the fairly conservative assumption that the relevant physics is
determined locally from the geometry of the singularities.

Finally in section \ref{s:num} we emphasize the peculiar numerical
predictions which arise from anomaly cancellation in the F-theory
compactification on $X$.

%%%%%%%%%%%%%%%%%%%%%%%%%%%%%%%%%%%%%%%%%%%%%%%%%%%%%%%%%%%%%%%%

\section{Lie algebras and \CY\ threefolds} \label{s:lcy}

We begin with
a \CY\ threefold $X$ which admits an elliptic fibration
$\pi:X\to\Sigma$, where $\Sigma$ is a complex surface, and also assume that
this
elliptic fibration has a section.\footnote{The F-theory limit cannot be
taken unless either the fibration has a section, or a $B$-field has been
turned on in the base \cite{BPS:Fquan,BKMT:IIBv,Mor:Hlect99}.} The type IIA string
compactified on $X$ 
yields an effective four-dimensional theory with $N=2$ supersymmetry; its
strong-coupling limit, known as ``M-theory compactified on $X$,'' yields an
effective five-dimensional theory.  One more effective spatial dimension is
obtained in a limit in which the areas of all components of the elliptic
fibers shrink to zero---this is the ``F-theory limit.''
See, for example, \cite{MV:F,me:lK3} for an explanation of this.

We point out that most of the following analysis does not really
depend upon this elliptic fibration structure and applies to M-theory
and type IIA compactifications of $X$. We use the F-theory language as
an organizational tool to give examples later on. One also has the
advantage in F-theory of being able to use anomaly cancellation as a
powerful tool in checking the consistency of results concerning
spectra of massless particles. In the F-theory context we can freely
exchange the notion 
of, say, an $\mathrm{I}_n$ fiber and an $A_{n-1}$ singularity. The
former is the elliptic fibration description for the latter. Recall
\cite{MV:F2} that this is because although $\mathrm{I}_n$ is really
the {\em extended\/} Dynkin diagram of $A_{n-1}$ and that one always ignores
the components of the fiber which hit the chosen section of the elliptic
fibration.\footnote{In fact, the F-theory limit should really be taken
in two steps: First, shrink to zero area
 all fiber components not meeting the chosen section, 
producing M-theory or the type IIA string compactified on a space with
ADE singularities; then shrink the remaining component of each
fiber down to zero area.} Thus, in the zero-area
fiber limit of F-theory, a shrunken $\mathrm{I}_n$ fiber 
gives the same physics as an $A_{n-1}$ singularity one dimension lower.

Whenever rational curves in $X$ are shrunk down to zero size we expect
$2$-branes of the type IIA string wrapped around these curves to
contribute massless particles to 
the spectrum. It is precisely these massless states which are the
focus of our interest in this paper.

Actually we need to be careful with the statement that massless states appear
automatically when a brane wraps a vanishing cycle. There is always
the subtlety of $B$-fields and R-R fields which should be tuned to the
right value (usually denoted ``zero'' by convention) to really obtain
a massless state. As emphasized in \cite{AD:tang} the relevant
parameters to worry about in this context are the R-R fields. 
We may see this as follows.
If one considers the type IIA string compactified on $X$ then 
deformations of the K\"ahler form (and $B$-field) on $X$ are given by
vector moduli. Suppose we use these K\"ahler moduli to
shrink down a holomorphic $2$-cycle to obtain an enhanced
gauge symmetry. Once we reach this point of
enhanced symmetry we may have a phase transition releasing new
hypermultiplet degrees of freedom. Thus at the point of phase
transition, these new parameters, which include R-R fields, are fixed
at some value.
Reversing this point of view, we
may tune parameters in the hypermultiplet moduli space to achieve an
enhanced gauge symmetry but these parameters include R-R fields. Thus
we need to assume always that the R-R parameters have been tuned to
the appropriate values required to obtain the enhanced gauge
symmetries we discuss below.

Witten \cite{W:MF} analyzed how to determine the massless particle
content for a given configuration of rational curves. Let us assume
that a given rational curve lives in a family parametrized by a moduli
space $M$. In the simplest case one has an embedding
$M\times\P^1\subset X$. An isolated rational curve is a trivial
example of this where $M$ is simply a point.

According to Witten's calculation,
 one half-hypermultiplet may be associated to the fact that a
$2$-brane breaks half of the supersymmetry. This half-hypermultiplet is
then tensored with the total cohomology of $M$
in an appropriate sense. The result is that if $M$
is a point, then we simply obtain a single half-hypermultiplet. If $M$
is an algebraic curve of genus $g$ then we obtain a single vector
multiplet and $g$ hypermultiplets. This was also argued by a different
method in \cite{KMP:enhg}. Note that for any wrapping we may also wrap
with the opposite orientation to double this spectrum.

Of central interest to us is the fact that compactifying a type IIA
string theory (and thus M-theory and F-theory) on a \CY\ threefold $X$ 
produces a 
theory with a Yang--Mills sector.  The gauge fields
 may be viewed as arising from integrating the R-R $3$-form of the
type IIA string over $2$-cycles in $X$ to produce 
$1$-forms.\footnote{In M-theory, one
likewise integrates the M-theory $3$-form field over $2$-cycles.}  These $1$-forms
play the role of the Yang-Mills connection. In addition, the $4$-cycles in
$X$ which are dual (via intersection theory) to these $2$-cycles will
play an important role.
Let $F$ denote the $4$-form field strength of the R-R $3$-form in the type
IIA string.
Note that the $2$-branes of the type IIA theory are {\em
electrically\/} charged under this field---that is 
\begin{equation}
  \int_{M_6} *F =1,  \label{eq:d}
\end{equation}
for a $6$-dimensional shape $M_6$ (such as a six-sphere) 
enclosing the seven directions
transverse to 
a fundamental $2$-brane.

Upon compactification we will be wrapping $2$-branes around a $2$-cycle in
$X$ to produce a point particle in four-dimensional space-time. To find the
charge of this resulting particle we may take $M_6=S^2\times S_i$,
where $S^2$ is a sphere in four dimensional space-time enclosing the
particle and $S_i$ is a $4$-cycle within $X$.

It follows that in the type IIA compactification
\begin{enumerate}
 \item We have $b_2(X)=b_4(X)$ gauge symmetries of the type
 $\mathrm{U(1)}$, 
 each labelled by an element of $H_4(X)$, in addition to the $U(1)$ gauge
 symmetry coming from the R-R $1$-form (whose charge is measured using
 $M_6=X$, the generator of $H_6(X)$).
 \item If a $2$-brane wraps a $2$-cycle $C_a$ to produce a particle then
 the ``electric'' charge of this particle under the $\mathrm{U(1)}$ symmetry
 associated to a $4$-cycle $S_i$ will be the intersection number
 $(S_i\cap C_a)$.
\end{enumerate}
We thus obtain a perturbative $\mathrm{U(1)}^{b_4(X)+1}$ gauge symmetry in type
IIA.  In 
the M-theory compactification, there is no R-R $1$-form, and the eight
transverse directions to the M-theory $2$-brane are enclosed by
$M_7=S^3\times S_i$, so the total 
``perturbative'' gauge symmetry is given by $\mathrm{U(1)}^{b_4(X)}$.  In the
F-theory limit, 
the only $4$-cycles which contribute gauge fields are those with
intersection number zero 
with the elliptic fiber; moreover, $4$-cycles which are the inverse images
of $2$-cycles in the base $\Sigma$ are associated to tensor multiplets
rather than gauge fields.  Thus, in the F-theory limit, we get a
``perturbative'' gauge symmetry group of 
$\mathrm{U(1)}^{b_4(X)-b_2(\Sigma)-1}$. 

As is now well-known, and as we will discuss, the wrapped $2$-branes will
elevate 
this $\mathrm{U(1)}^{b_4(X)+\varepsilon}$ gauge symmetry to a non-abelian Lie
group (since certain wrapped branes include vector multiplets in their
spectra), where $\varepsilon=1$, $0$, or $-b_2(\Sigma)-1$ for IIA, M-theory
or F-theory, respectively. From
now on we will concern ourselves only with the Lie {\em 
algebra\/} of the gauge symmetry. It was noted in \cite{AM:frac} that, at
least in F-theory,
the global structure of the gauge group may be recovered from the
Mordell-Weil group of $X$ as an elliptic fibration.\footnote{Indeed
$\pi_1$ of the gauge group is equal to the Mordell-Weil group
(including both the free and torsion parts).} 

If this $\gu(1)^{\oplus (b_4(X)+\varepsilon)}$ appears as the Cartan
subalgebra of  
our gauge algebra then the discussion above implies that we may 
make the following identifications. Let $\mathfrak{h}$ be
the (real) Cartan subalgebra, let $\mathfrak{h}^*$ be the
dual space, and let $\Lambda\subset\mathfrak{h}$ be the coroot lattice and
$\Lambda^*\subset\mathfrak{h}^*$ be the weight lattice so that 
the Cartan subgroup $\mathrm{U(1)}^{b_4(X)+\varepsilon}$ is naturally
identified with 
$\mathfrak{h}/\Lambda$.  For the IIA compactification, we take
$\Lambda=H_4(X,\mathbb{Z})\oplus H_6(X,\mathbb{Z})$
and $\Lambda^*=H_0(X,\mathbb{Z})\oplus H_2(X,\mathbb{Z})$, and in M-theory,
we take 
$\Lambda=H_4(X,\mathbb{Z})$ and $\Lambda^*=H_2(X,\mathbb{Z})$.
In F-theory, we begin with the orthogonal complement within
$H_4(X)$ of the elliptic fiber $E$, and then we mod out by
$\pi^{-1}H_2(\Sigma)$  (that is,
$\Lambda=[E]^\perp/\pi^{-1}H_2(\Sigma)\subset H_4(X)/\pi^{-1}H_2(\Sigma)$);
we then take  
$\Lambda^*=\Hom(\Lambda,\mathbb{Z})$ to be the dual 
lattice of $\Lambda$. 
In each case, a
$2$-brane wrapped around a particular $2$-cycle is then naturally
associated with an element of the 
weight lattice
and its charges under
the Cartan subalgebra are given in the standard way.

We work this out in detail in several particular cases.
Consider first the case that $X$ contains a ``ruled'' complex surface
$S$ admitting a  
fibration $\pi:S\to M$, for some $M$, where all fibers are isomorphic
to $\P^1$. 
The fibers will shrink down to zero size in the F-theory limit.
The simplest example of this is $M\times C_1$ where $M$ is
a Riemann surface of genus $g$ and that $C_1\cong\P^1$ is in the
fiber direction.
That is to say, in our elliptic fibration $\pi:X\to\Sigma$ we have a 
curve $M\subset\Sigma$ over which the fiber is
I$_2$. Clearly we have massless states appearing for the $2$-branes
wrapped around $C_1$. We also have a $\gu(1)$ symmetry associated to 
$S_1\cong M\times C_1$. Let us consider the normal bundle of a single
$C_1$ curve. This normal bundle may be written as $\O(a)\oplus\O(b)$
where $a+b=-2$ by the adjunction formula and the fact that $X$ is a
\CY\ space. Since this curve may be translated along the $M$ direction
one of these line bundles must be trivial. Thus the normal bundle is
$\O\oplus\O(-2)$ where the $\O(-2)$ describes the normal bundle
direction which is also normal to $S_1$. Therefore $(S_1\cap
C_1)=-2$. This tells us that we have a vector supermultiplet and $g$ 
hypermultiplets from wrapping $2$-branes around $M\times\P^1$ all with
charge $-2$ with respect to the $\gu(1)$ gauge symmetry associated to
this divisor. Similarly by wrapping with the opposite orientation we
obtain a copy of this except with charge $+2$.

These vector supermultiplets enhance the $\gu(1)$ symmetry to $\su(2)$
in the usual way and we have an additional $g$ hypermultiplets in the
adjoint representation. The key point is to notice in this
construction that the condition
\begin{equation}
  (S_1\cap C_1)=-2,
\end{equation}
has played the role of the {\em Cartan matrix\/} of 
$\su(2)$.\footnote{Of course there is a sign difference here compared
to usual Lie algebra theory. This sign difference is purely due to the
convention that Lie algebra theorists insist on the Cartan matrix
being positive definite, rather than negative definite. If string
theory had been studied before Lie algebras then the sign would be the
other way!}

The next simplest case is where we have a set of curves
$C_1,\ldots,C_n$ which may intersect each other and are each
isomorphic to $\P^1$ and lying in the fiber direction. We assume that
$M\times (\bigcup_i C_i)$ embeds algebraically into $X$.\footnote{We can
consider more generally a situation where we glue together $n$ 
distinct $\P^1$ fibrations over $M$ along appropriate disjoint sections,
forming a chain.  In the remainder of this paper, we will continue
to explain by example and will not explicitly state the most general
form of the algebraic surfaces which contract to $M$ in the F-theory limit.}
We now have a
Cartan matrix given purely by the configuration of
$C_1,\ldots,C_n$. Applying the above method we obtain the standard
F-theory result of a simply-laced enhanced gauge symmetry
as listed, for example, in table 4 of \cite{me:lK3}.

\iffigs
\begin{figure}
\begin{center}
\setlength{\unitlength}{0.00041700in}%
\begin{picture}(8865,6298)(511,-6347)
\thinlines
\put(1201,-1861){\line( 5, 2){4500}}
\put(1201,-1861){\line( 0,-1){3900}}
\put(1201,-5761){\line( 5, 2){4500}}
\put(5701,-61){\line( 0,-1){3900}}
\put(3631,-511){\line( 2,-1){4350}}
\put(1984,-1180){\line( 2,-1){4350}}
\put(1201,-2774){\line( 5, 2){4500}}
\put(1201,-3689){\line( 5, 2){4500}}
\put(1201,-4591){\line( 5, 2){4500}}
\put(1201,-5253){\line( 5, 2){4500}}
\put(3619,-1428){\line( 2,-1){4350}}
\put(3579,-2326){\line( 2,-1){4350}}
\put(3469,-3190){\line( 2,-1){4350}}
\put(3290,-4023){\line( 2,-1){4350}}
\put(2692,-3462){\line( 2,-1){4350}}
\put(2272,-2172){\line( 2,-1){4350}}
\put(8401,-1861){\vector(-1,-1){1275}}
\put(9301,-3961){\vector(-1, 0){1275}}
\put(3301,-5761){\vector(-3, 4){1125}}
\put(6901,-1261){\vector(-1, 0){1125}}
\put(3081,-4165){\line( 2,-1){4350}}
\put(601,-5161){\vector( 0, 1){2955}}
\put(2401,-1381){
}
\put(1973,-1178){
}
\put(6332,-3356){
}
\put(8476,-1936){\makebox(0,0)[lb]{\smash{$C_2$}}}
\put(9376,-4036){\makebox(0,0)[lb]{\smash{$S_2$}}}
\put(6976,-1336){\makebox(0,0)[lb]{\smash{$S_1$}}}
\put(3376,-5911){\makebox(0,0)[lb]{\smash{$C_1$}}}
\put(411,-5561){\makebox(0,0)[lb]{\smash{$M_1$}}}
\end{picture}
\end{center}
  \caption{Ruled surfaces producing $\sp(2)$.}
  \label{fig:one}
\end{figure}
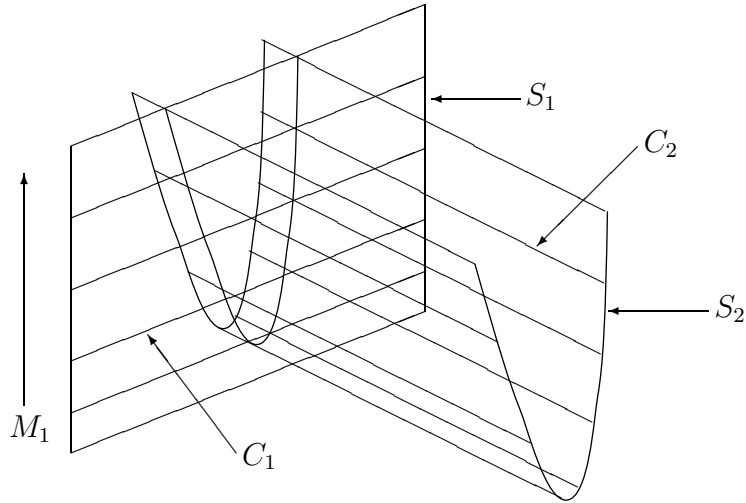
\fi

As noted first in \cite{IMS:5deg} the real power of this Cartan matrix
approach is that it gives a clear way of describing {\em
non}-simply-laced gauge algebras. Consider a less trivial example of
ruled surfaces as shown in Figure~\ref{fig:one}. In this example the
moduli space $M_1$ of the curve $C_1$ is different from the moduli space
$M_2$ the curve $C_2$. 
Think of $M_1$ as the vertical direction in the figure.
We obtain ruled surfaces $S_1=M_1\times C_1$
and $S_2=M_2\times C_2$. We have a two-fold cover $M_2\to M_1$ branched
at one point in Figure~\ref{fig:one}. Any other branch points are not
shown in the figure.

\def\fkm{\phantom{-}}
The intersection matrix of this configuration may be written
\begin{equation}
  (S_i\cap C_j) = \begin{pmatrix} -2&\fkm1\\\fkm2&-2\end{pmatrix}.
\end{equation}
This is the Cartan matrix for $\sp(2)$ (or $\so(5)$) and so our
enhanced gauge algebra should be $\sp(2)$.

This phenomenon of obtaining a non-simply-laced symmetry algebra was
first noted in \cite{AG:sp32} inspired by the construction of
\cite{CHL:bigN}. There it was explained by monodromy acting on the
fibers as follows. Let $M_1$ be embedded in $\Sigma$ and let $M_2$ be
a two-fold cover of $M_1$ (branched at various points). Over a generic
point in $M_1$ we see that, ignoring the component meeting the chosen
section, the 
Kodaira fiber consists of one line from $S_1$ and two lines from $S_2$
forming the (dual) Dynkin diagram of $\su(4)$. Moving along a closed path
in the complement of the set of branch points of $M_2\to M_1$ we will
exchange the two lines in $S_2$. This action on the Dynkin diagram
is induced by an {\em outer automorphism\/} of $\su(4)$ and the
invariant subgroup under this outer automorphism can be taken to be
$\sp(2)$.

One might therefore suspect that the effective gauge algebra
is the monodromy-invariant subalgebra of the simply-laced gauge
symmetry generated locally by the vanishing cycles. This was the
assertion in \cite{AG:sp32}. Unfortunately it is an ambiguous
statement.\footnote{It should be possible to resolve this ambiguity by 
exhibiting the gauge algebra structure itself
(and not just the Cartan matrix) 
along the lines of \cite{HM:alg2}.  We leave this for future work, and in this
paper we shall resort to less direct arguments.}

Let us analyze carefully all possible outer automorphisms of
$\SU(2k)$. An element $g$ of $\SU(2k)$ satisfies $({}^T\bar{g})g=1$.
Complex conjugation $t:g\mapsto\bar{g}$ is an example of an outer
automorphism.  Indeed this acts on the Dynkin diagram of $\SU(2k)$ by
reflection about the middle node. Clearly the invariant subgroup under
this outer automorphism is given by $g$ real. But this yields the
group $\SO(2k)$---not what we were expecting!

A general outer automorphism of $\SU(2k)$ can be obtained by combining
complex conjugation with an
arbitrary inner automorphism, yielding $g\mapsto h^{-1}\bar{g}h$, where
$h\in\SU(2k)$ (there are no
other possibilities since that would imply further symmetries of the
Dynkin diagram). Since this outer automorphism acts on the Dynkin diagram
as the reflection, it is also a viable candidate for the monodromy action
on the gauge group.  In this general situation, the invariant subalgebra 
satisfies
\begin{equation}
  ({}^Tg)hg=h.   \label{eq:inv}
\end{equation}
The case $h=1$ yields $\SO(2k)$ as stated before. Now if we put
\begin{equation}
  h = \begin{pmatrix} \fkm0&I\\ -I&0\end{pmatrix},
\end{equation}
where $I$ is the $k\times k$ identity matrix, we obtain the group
$\Sp(k)$---as desired.  In this case, the outer automorphism is an involution,
but this is not a requirement in general.

We see then that the method of directly working out the Cartan matrix
from intersection theory is a better way to determine the effective
gauge algebra in F-theory
than trying to find subalgebras invariant under outer
automorphisms. The latter method is ambiguous. One might try to assert
that F-theory picks out the ``maximal'' invariant subalgebra under all
possible outer automorphisms. Indeed, $\sp(k)$ is ``bigger'' than
$\so(2k)$ in as much as it has a larger dimension (although in general
$\so(2k)\not\subset\sp(k)$). However, even this approach is 
inadequate
as will be shown by examples in section \ref{s:odd}.

Note that in the M-theory or type IIA compactifications, an ambiguity of
the sort we have discovered is actually to be expected.  As we have already
pointed out, if the Ramond--Ramond fields have non-zero expectation values
then some of the non-abelian gauge fields will become massive;  when
these are integrated out, the gauge group becomes smaller.  This is
precisely what happens when the outer automorphism of the covering Lie
group is varied in the construction above.  The gauge algebra which we wish
to determine is the one in which these effects have been turned off so that
the F-theory limit can be taken.  (A similar phenomenon of variable gauge
group depending on the precise value of an outer automorphism has been
observed in a closely related context by Witten \cite{W:N6d}, and applied
in \cite{BCD:,BL:}).

%%%%%%%%%%%%%%%%%%%%%%%%%%%%%%%%%%%%%%%%%%%%%%%%%%%%%%%%%%%%%%%%

\section{Counting hypermultiplets}  \label{s:hyper}

In the last section we described how to determine the gauge algebra in
F-theory (or M-theory or IIA string theory) by determining the Cartan
matrix from intersection theory. Similar methods will in principle
determine the 
hypermultiplet spectrum completely as we now discuss.

First there can be the case of a family of rational curves acquiring
extra rational curves at certain points in the family. In the context
of elliptic fibrations this can be seen as collisions of curves in
$\Sigma$ over which there are singular fibers. The simplest example is
a transverse collision of I$_n$ and I$_m$. 

The resolution of singularities associated to
this collision was explained in \cite{Mir:fibr}, and applied to the
case of an I$_n$-I$_1$ collision in the
context of string theory in section 8.2 of \cite{IMS:5deg}. The key
point is that there exist rational curves within the collision with
normal bundle $\O(-1)\oplus\O(-1)$. One of these curves $C$ is the 
intersection of two ruled $4$-cycles, one lying over the curve of $I_n$
fibers, and the other lying over the curve of $I_m$ fibers.  The 
normal bundle of $C$ in $X$ is naturally the direct sum of the
normal bundles of $C$ in each of these $4$-cycles, and each of these is
$\O(-1)$. Thus this
curve appears as (minus) a fundamental weight. 
The above rules imply that we
have found a curve representing the (lowest) weight of the
$(\mathbf{n},\mathbf{m})$ representation of $\su(n)\oplus\su(m)$. By
adding other (possibly reducible) curves in the collision fiber we may
indeed build up the full $(\mathbf{n},\mathbf{m})$ representation.
Thus {\em the transverse collision of a curve of I$_n$ and I$_m$
  fibers yields a hypermultiplet in the $(\mathbf{n},\mathbf{m})$
  representation of $\su(n)\oplus\su(m)$.} (This same result had earlier
been determined using quite different methods in \cite{BSV:D-man}.  Another
approach which is closer to ours appeared in \cite{KV:hyp}.)

Many other ``collisions'' can be explained in similar ways.  However, if
the extra rational curves at the collision point have normal bundles other
than $\O(-1)\oplus\O(-1)$, then Witten's calculation does not directly
apply.  General methods for evaluating the corresponding contribution to
the hypermultiplet spectrum are not known.

The case of non-simply-laced symmetry algebras raises even more complicated
possibilities. Some of the hypermultiplet matter can appear in a
somewhat ``non-local'' manner as we now explain. Suppose we are in a
situation analogous to Figure~\ref{fig:one}. Let us consider the
example of a type $\mathrm{I}_{2k}$ fiber (where we again ignore the component
passing through the chosen section). Let the middle component in the chain
have a moduli space given by $M_1$ and the other components have
moduli space $M_2$ where $M_2\to M_1$ is a double cover. That is, we
have a $\Z_2$-monodromy acting on the $\mathrm{I}_{2k}$ fiber (in the only
possible way). Figure~\ref{fig:one} is the case $k=2$.

According to \cite{W:MF} we should obtain $g(M_1)$ hypermultiplets for
$2$-branes wrapping the middle component and $g(M_2)$ hypermultiplets for
$2$-branes wrapping each of the other components. Note that
$g(M_2)\geq g(M_1)$ from the double cover. There are additional 
hypermultiplets arising from wrapping connected unions of these components.
In fact, each of the positive roots of the covering algebra $\su(2k)$ is
represented 
by such a connected union, some of which are fixed by the monodromy, and
others of which are exchanged in pairs under the monodromy.  The ones which
are fixed under monodromy have $M_1$ as moduli space, while those which are
exchanged in pairs have $M_2$ as their moduli space.

When we organize these weights in terms of representations of $\sp(k)$, we
find that the invariant subspace describes the adjoint of $\sp(k)$ while
the anti-invariant subspace describes the remaining weights in the adjoint
of $\su(2k)$.  On the other hand, each invariant positive root contributes
to the invariant subspace, while the roots exchanged in pairs contribute to
both the invariant and anti-invariant subspaces.  We conclude that the
adjoint of $\sp(k)$ occurs $g(M_1)$ times while the weights in the
anti-invariant subspace each occur $g(M_2)-g(M_1)$ times.

\iffigs
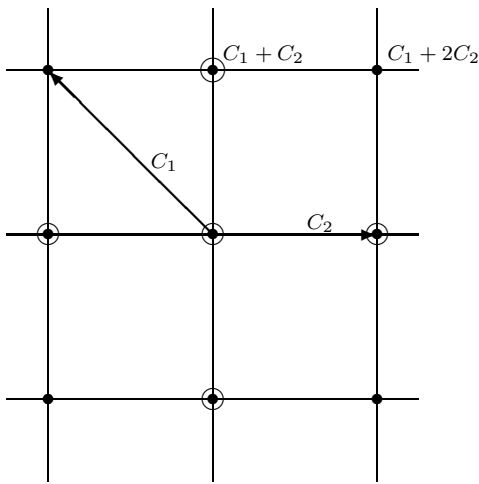
\begin{figure}
\begin{center}
\setlength{\unitlength}{1700sp}%
\begin{picture}(6024,6924)(2389,-6673)
\thinlines
\put(3001,-661){\circle*{150}}
\put(5401,-661){\circle*{150}}
\put(7801,-661){\circle*{150}}
\put(3001,-3061){\circle*{150}}
\put(7801,-3061){\circle*{150}}
\put(3001,-5461){\circle*{150}}
\put(5401,-5461){\circle*{150}}
\put(7801,-5461){\circle*{150}}
\put(5401,-3061){\circle*{150}}
\put(5401,-661){\circle{336}}
\put(3001,-3061){\circle{300}}
\put(5401,-3061){\circle{300}}
\put(7801,-3061){\circle{300}}
\put(5401,-5461){\circle{300}}
\put(5401,239){\line( 0,-1){6900}}
\put(7801,239){\line( 0,-1){6900}}
\put(3001,239){\line( 0,-1){6900}}
\put(2401,-3061){\line( 1, 0){6000}}
\put(2401,-5461){\line( 1, 0){6000}}
\put(2401,-661){\line( 1, 0){6000}}
\thicklines
\put(5401,-3061){\vector( 1, 0){2400}}
\put(5401,-3061){\vector(-1, 1){2400}}
\put(6776,-2986){\makebox(0,0)[lb]{\smash{\scriptsize $C_2$}}}
\put(4501,-2086){\makebox(0,0)[lb]{\smash{\scriptsize $C_1$}}}
\put(5551,-511){\makebox(0,0)[lb]{\smash{\scriptsize $C_1+C_2$}}}
\put(7951,-511){\makebox(0,0)[lb]{\smash{\scriptsize $C_1+2C_2$}}}
\end{picture}
\end{center}
  \caption{The adjoint of $\sp(2)$.}
  \label{fig:asym}
\end{figure}
\fi

We demonstrate which weights appear in the example of $k=2$ in 
Figure~\ref{fig:asym}. We show the weights of the adjoint representation.
The dots represent weights associated to $M_1$,
i.e., from $C_1$. The circles represent weights associated to
$M_2$, i.e., from $C_2$. 
It is important to note that {\em reducible\/} curves may also be
wrapped by $2$-branes. That is, two rational curves intersecting
transversely at a point may be viewed together as a nodal rational
curve. These wrappings of reducible curves are required to obtain all
the adjoint weights of the vector multiplets of the previous section.
The reducible curve $C_1+C_2$ has moduli space given by
$M_2$---since $C_2$ has a moduli space given by $M_2$. Looking at
Figure~\ref{fig:one} we see that there is also a chain of rational
curves in the class $C_1+2C_2$ but note this this combination is
invariant under the $\Z_2$ monodromy and so has moduli space given by
$M_1$. 
These circles form the weights of $\mathbf{5}$ of $\sp(2)$.
Ignoring the zero weights for now we see that the adjoint
appears $g(M_1)$ times and the $\mathbf{5}$ of $\sp(2)$ appears
$g(M_2)-g(M_1)$ times.

Indeed the zero weights also work out correctly. A zero weight must
represent an uncharged hypermultiplet and therefore a modulus. We may
use the work of Wilson \cite{Wil:Kc} to demonstrate this. Wilson showed
that a \CY\ threefold containing a ruled surface $M\times\P^1$ has a
moduli space which preserves this ruled surface only in codimension
$g(M)$. That is, there are $g(M)$ deformations of the \CY\ threefold
which destroy this ruled surface.  Applying this to both ruled
surfaces, we get $g(M_1)+g(M_2)$ deformations.  On the other hand,
each $\sp(2)$ adjoint contains a two-dimensional weight zero eigenspace 
while each $\mathbf{5}$ contains a one-dimensional weight zero eigenspace.
Thus the dimension of the weight zero eigenspace is $2(g(M_1))+
(g(M_2)-g(M_1))$, which simplifies to $g(M_1)+g(M_2)$, as claimed.

The above construction may be easily generalized to $\sp(k)$:\footnote{The
explanation given here was applied in \cite{IMS:5deg} to obtain a detailed 
picture of the surfaces which collapse as the gauge symmetry is enhanced.}
\begin{theorem}
\label{ieven}
  Let $\Z_2$ monodromy act on an I$_{2k}$ fiber in F-theory so that the
  central component of the fiber has moduli space $M_1$ and the outer
  components have moduli space $M_2$. Thus $M_2\to M_1$ is a double
  cover. Then the resulting gauge algebra is $\sp(k)$ and we have
  $g(M_1)$ hypermultiplets in the adjoint representation and 
  $g(M_2)-g(M_1)$ hypermultiplets in the $\Lambda^2$ representation
  (which has dimension $k(2k-1)-1$).
\end{theorem}

Similarly $\mf{e}_6$ with $\Z_2$ monodromy will yield an $\mf{f}_4$
gauge algebra with $g(M_2)-g(M_1)$ hypermultiplets in the
$\mathbf{26}$ representation (in addition to the usual $g(M_1)$
adjoints). Also $\so(2k)$ with $\Z_2$ monodromy will yield an
$\so(2k-1)$ gauge algebra with $g(M_2)-g(M_1)$ hypermultiplets in the
vector $\mathbf{2k-1}$ representation. In the case of $\so(8)$ with $\Z_3$
or $\mathfrak{S}_3$ monodromy, a similar analysis yields $g(M_2)-g(M_1)$
hypermultiplets in the $\mathbf{7}$ representation of $\mf{g}_2$.

This agrees with the various computations in \cite{BKV:enhg} where
$M_1\cong\P^1$. Let $M_2$ be the double cover of $M_1$ branched at $b$
points. Thus 
$g(M_2)=\ff12b-1$. Then, for example, in the $\mf{f}_4$ case we should
have $\ff12b-1$ $\mathbf{26}$'s. This agrees with section 4.3 of
\cite{BKV:enhg} by identifying the branch points with the $b=2n+12$
zeroes of $g_{2n+12}$.

One might note that the above cases with $\Z_2$ monodromy may be
combined into a simple rule 
as follows. Let $\mf{s}$ be the simply-laced Lie algebra which
contains the actual gauge symmetry algebra $\mf{g}$ as a subalgebra
invariant under an outer automorphism given by the monodromy
action. 
(In fact, in each of the above cases, the outer automorphism which we use
actually has order $2$ as an automorphism of $\mf{s}$, not merely order $2$
as an automorphism of the Dynkin diagram.)
We may then decompose the adjoint representation of
$\mf{s}$ as follows
\begin{equation}
  \Ad(\mf{s}) = \Ad(\mf{g})\oplus V_{-}   \label{eq:rul1}
\end{equation}
where $V_{-}$ is a (possibly reducible) representation of
$\mf{g}$ on which the generator of the $\Z_2$ outer automorphism acts
as $-1$. The above rules may be combined to say that we obtain
$g(M_2)-g(M_1)$ hypermultiplets in the $V_{-}$ representation. 
As we have already noted above in the case $\mf{g}=\mf{g}_2$, the rule will
be different if the monodromy group is not $\Z_2$.
In fact, we will see more generally in section \ref{ss:small} that 
if the outer automorphism representing the monodromy has higher order,  the
simple rule expressed in equation~\eqref{eq:rul1} must be modified.  

In addition to these ``non-local'' hypermultiplets coming from
rational curves moving in families one may also obtain further
hypermultiplets from collisions of curves of reducible fibers
as in the I$_n$-I$_m$ collision discussed above.\footnote{An argument for
why certain hypermultiplets appear to be 
``local''---i.e., tied to isolated rational curves---or ``non-local''
was given in \cite{FMW:F}.}
Note that some simple
collisions may just induce monodromy without further contributions
(that is, their contributions are completely accounted for by the
representation $V_-$ obtained in eq.~(\ref{eq:rul1}) ). As
an example we show in Figure~\ref{fig:so9} the generic
case of a Spin(9) gauge symmetry in F-theory.

\iffigs
\begin{figure}
\begin{center}
\setlength{\unitlength}{0.00050000in}%
\begin{picture}(7512,3045)(2389,-2803)
\thinlines
\put(2401,-1561){\line( 1, 0){6000}}
\put(6901,-61){\line( 0,-1){2700}}
\put(3301,-2761){
}
\put(8551,-1636){\makebox(0,0)[lb]{\smash{$\Ist1$}}}
\put(9901,-1561){\makebox(0,0)[lb]{\smash{$\HS{-n}$}}}
\put(3451,-2761){\makebox(0,0)[lb]{\smash{$\mathrm{I}_1$}}}
\put(7051,-2761){\makebox(0,0)[lb]{\smash{$\mathrm{I}_1$}}}
\put(4501, 14){\makebox(0,0)[lb]{\smash{$\times (n+4)$}}}
\put(6751, 14){\makebox(0,0)[lb]{\smash{$\times 2(n+6)$}}}
\end{picture}
\end{center}
  \caption{The generic case of Spin(9) gauge symmetry.}
  \label{fig:so9}
\end{figure}
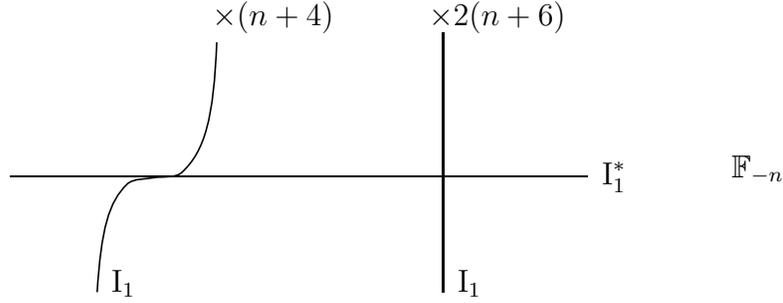
\fi

This figure shows an $\Ist1$ fiber along a section of the Hirzebruch
surface $\HS{n}$. This section has self-intersection $+n$ and
is denoted 
$C_0$ in the notation of \cite{me:lK3}. In the most generic situation,
the rest of the discriminant locus of the elliptic fibration will
consist of $\mathrm{I}_1$ fibers along curves which intersect $C_0$ as
shown in Figure~\ref{fig:so9}. 
Generically there are two types of collisions occurring with the
frequencies shown.
A lengthy computation shows that the
$n+4$ cubic collisions\footnote{Locally these cubic collisions may be
written in Weierstrass form 
as $y^2 = x^3-3s^2t^2x+2s^3(s+t^3)$ where $s$ and $t$ are affine
coordinates in $\HS{-n}$.}
produce extra rational curves in the fiber but
no monodromy while the $2(n+6)$ transverse collisions produce monodromy
but no extra rational curves. Thus the $2(n+6)$ collisions produce
$n+5$ of the vector $\mathbf{9}$'s of $\so(9)$. An analysis of the
rational curves in the cubic collision shows that we have $n+4$ spinor
$\mathbf{16}$'s. Assuming $n\neq-4$, the existence of these spinors
shows that the gauge group must be Spin(9). This agrees perfectly with
section 4.6 of \cite{BKV:enhg}. Similarly all the other results of
\cite{BKV:enhg} may be confirmed.

Finally in this section let us return to the case of $\Z_2$ monodromy
acting on a curve of $\su(2k)$ singularities to give an effective
$\sp(k)$ gauge symmetry. 
We will consider the Higgs branch in which we give expectation values
to the hypermultiplets so as to break completely this $\sp(k)$ gauge
symmetry. 
Recall that the geometry of moduli spaces of supersymmetric field
theories in question imply that the dimension of this Higgs branch
should equal the total dimension of the representations of charged
hypermultiplets minus the dimension of the gauge group which is broken.
We will observe that the
geometry is in accord with Theorem \ref{ieven}.  We do this by
describing the deformations after shrinking all of the curves in the
fibers to zero volume.  In section \ref{ss:below} we will use the
ideas introduced here towards the justification of 
our Main Assertion
stated in the next section, which 
states
that the gauge
algebra in the case of $\su(2k+1)$ with $\Z_2$ monodromy is $\sp(k)$.

We let $\pi:M_2\to M_1$ be an unramified (for
simplicity) double cover of $M_1$. In addition, we denote by
$\iota:M_2\to M_2$ the involution which exchanges sheets of the double
cover.  We now describe a local \CY\ threefold $X$ containing the
geometry of $\su(2k)$ with $\Z_2$ monodromy over $M_1$.  First we
construct a \CY\ threefold $Y$ with an $\su(2k)$ fibration over $M_2$
without monodromy.  Then $X$ will be constructed as a $\Z_2$ quotient of 
$Y$.

We construct a singular threefold inside the
bundle $V=K_{M_2}^k\oplus 
K_{M_2}^k \oplus K_{M_2}$ as the variety defined by the
equation
\begin{equation}
xy=z^{2k}
\label{eq:su2k}
\end{equation}
where $x$ and $y$ are in $K_{M_2}^k$ and $z$ is in $K_{M_2}$.  This
threefold has an $A_{2k-1}$ singularity along $M_2$, which is
identified with the zero section of $V$.  It has trivial
canonical bundle by the adjunction formula.  In a moment we will
construct a nowhere vanishing holomorphic $3$-form on it, giving
independent verification of this fact. The desired threefold $Y$ is
obtained by blowing up the singular locus $k$ times in the usual way
to obtain a chain of $2k-1$ ruled surfaces over $M_2$.

To obtain the desired geometry, we take the quotient of $Y$ by the
fixed point free
involution obtained by using $\iota$ on $M_2$ while sending $(x,y,z)$
to $(y,x,-z)$.  Note that the fibers of $K_{M_2}$ over $p\in M_2$ and
$\iota(p)\in M_2$ are canonically identified with the fiber of
$K_{M_1}$ over $\pi(p)$, so this map makes sense.  Using the explicit
description of the blowup and the fact that $x$ and $y$ are
interchanged, it follows that there is $\Z_2$ monodromy.

To show that the quotient $X$ by this involution has trivial canonical
bundle, it suffices to show that the involution preserves the
holomorphic $3$-form on $Y$.  It suffices for our purposes to compute on
the singular model.  Let $\omega$ be any holomorphic $1$-form on $M_2$.
Then
\begin{equation}
\omega\wedge dx \wedge dy\wedge dz
\end{equation}
is a holomorphic $4$-form on $V$
with values in $K_M^{2k+1}$.
Now, thinking of $\omega$ as a section of $K_{M_2}$, we divide by
$\omega$ to obtain the nowhere vanishing $4$-form
\begin{equation}
\left(\omega\wedge dx \wedge dy\wedge dz\right)/\omega
\end{equation}
on $V$ with values in $K_{M_2}^{2k}$ which is independent of $\omega$.
Finally, the residue
\begin{equation}
\mathrm{Res}\left(\frac{\left(\omega\wedge dx \wedge dy\wedge dz\right)/\omega}
{xy-z^{2k}}\right)
\end{equation}
is the holomorphic $3$-form on the singular model of $Y$.
It is clearly invariant under the involution. 

The deformations of $X$ may be described as the deformations of $Y$
in $V$ which preserve the involution.  The general deformation of
$Y$ (up to change of coordinates)
is given by 
\begin{equation}
xy=z^{2k}+ \sum_{i=2}^{2k} f_{i}z^{2k-i},
\label{eq:su2kdef}
\end{equation}
where the $f_i$ are sections of $K_{M_2}^i$.  Note that we are implicitly
assuming $g(M_2)>0$ to construct these deformations.  The invariance
condition is that $f_i$ lies in the $(-1)^i$-eigenspace of $\iota$.

We now count parameters.  The $+1$-eigenspace of $H^0(K_{M_2}^i)$ has
dimension $(2i-1)(g(M_1)-1)$, while the $-1$-eigenspace has dimension
$(2i-1)(g(M_2)-g(M_1))$.  Thus the dimension of the Higgs branch is
\begin{equation}
\begin{aligned}
\ &(3+7+\ldots + (4k-1))(g(M_1)-1)+(5+9+\ldots+(4k-3))(g(M_2)-g(M_1))\\
=&k(2k+1)(g(M_1)-1)+\left(k(2k-1)-1\right)(g(M_2)-g(M_1)).\\
\end{aligned}
\end{equation}
This is exactly the number of parameters freed up from the $g(M_1)$
adjoints and $g(M_2)-g(M_1)$ copies of the $\Lambda^2$ representation
by Higgsing an $\sp(k)$, as expected.  We have implicitly assumed that 
there are no global obstructions to the local deformations that we have
constructed above, and our parameter count is consistent with this 
assumption.

%%%%%%%%%%%%%%%%%%%%%%%%%%%%%%%%%%%%%%%%%%%%%%%%%%%%%%%%%%%%%%%%

\section{The case of monodromy on $\su(\textrm{odd})$.}  \label{s:odd}

While we appear to have given fairly general rules in the previous section for
computing the massless particle spectrum of an F-theory
compactification, there are actually many cases where the rules we
have given so far become difficult to apply.

In particular, Witten's analysis of the moduli space of rational
curves in \cite{W:MF} assumes that everything is smooth (and
reduced). This need not be the case. We will discuss some awkward
cases which appear quite commonly in F-theory.

We begin with a discussion of a case which has caused some confusion
in the literature---that of $\Z_2$ monodromy acting on a gauge
algebra of $\su(\mathrm{2k+1})$. The approach of asking for the
largest subalgebra invariant under an outer automorphism is not that
helpful in this
case. Putting $h=1$ in eq.~(\ref{eq:inv}) shows that $\so(\mathrm{2k+1})$
is one possibility. Decomposing $2k+1$ as $k+1+k$ and putting
\begin{equation}
 h = \begin{pmatrix} \fkm0&0&I\\ \fkm0&1&0\\ -I&0&0\end{pmatrix}, \label{eq:h} 
%h = \begin{pmatrix} \fkm0&I&0\\ -I&0&0\\ \fkm0&0&1\end{pmatrix}, \label{eq:h} 
\end{equation}
(where $I$ is the $k\times k$ identity matrix) shows that
$\sp(k)$ is another possibility. 
(This form of $h$ is nicely adapted to the action on the Dynkin diagram.) 
Now it so happens that 
$\dim(\so(2k+1))=\dim(\sp(k))$ (and that this is the largest dimension
which can occur). So which is the gauge algebra that
F-theory actually wants?

\iffigs
\begin{figure}
  \centerline{\epsfxsize=11cm\epsfbox{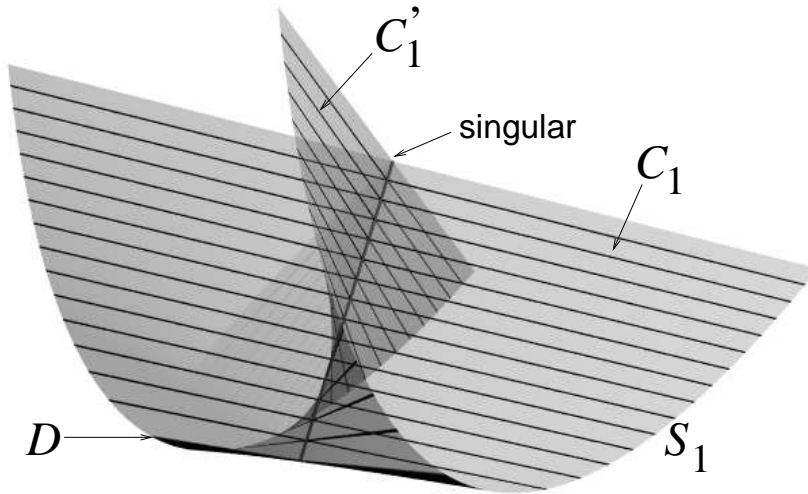}}
  \caption{The singular surface for monodromy in $\su(3)$.}
  \label{fig:nd}
\end{figure}
\fi

Using the approach of section \ref{s:lcy} we immediately run into a
problem. One of the ruled surfaces, which we will denote $S_1$, swept out by
the reducible components of the fibers will look inevitably locally
like the surface 
\begin{equation}
  y^2-x^2z=0  \label{eq:prob}
\end{equation}
in $\C^3$.
We show a sketch of (the real version of) this surface in 
Figure~\ref{fig:nd}. 
Each line $C_1$ in this surface crosses another line $C_1'$ in the
same class.
In the case of $\su(2k+1)$ for $k>1$ there will be other
smooth surfaces. This case is a little hard to visualize. In 
Figure~\ref{fig:nd2} we show the case of monodromy acting on $\su(5)$. 
(In this case $S_2$ is the surface $z=x^2$.) The
thick lines at the bottom of this sketch show the fiber over a branch point
of $M_2\to M_1$.

The problem is that it is not clear what value we should give to
$S_1\cap C_1$ since $C_1$ meets the singular line in $S_1$. 
The most na\"{\i}ve interpretation of Figure~\ref{fig:nd2} is to
completely ignore the fact that $S_1$ is singular and from the figure
read off the intersection matrix
\begin{equation}
  (S_i\cap C_j) = \begin{pmatrix} -2&\fkm1\\\fkm1&-2\end{pmatrix}.
\end{equation}
This would imply that the gauge algebra is $\su(3)$. In general, according
to this argument, $\Z_2$
monodromy acting on $\su(2k+1)$ would produce $\su(k+1)$---neither
of the possibilities suggested above! It would be the most obvious
algebra suggested by ``folding the Dynkin diagram up'' by the outer
automorphism.

\iffigs
\begin{figure}
  \centerline{\epsfxsize=11cm\epsfbox{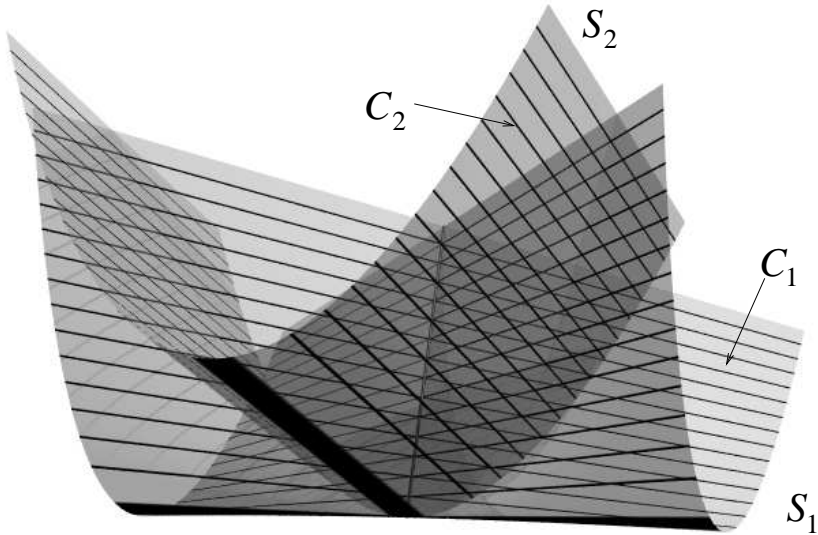}}
  \caption{The singular surface for monodromy in $\su(5)$.}
  \label{fig:nd2}
\end{figure}
\fi

We could get an $\sp(k)$ Cartan matrix from the case in question if we
could somehow tie $C_1$ and $C_1'$ in Figure~\ref{fig:nd}
together. That is, somehow the rules of $2$-brane wrapping would have
to assert that $C_1$ may not be wrapped alone---one must also wrap
the intersecting curve $C_1'$ simultaneously. Since $S_1$ is singular,
there is no known reason for ruling such a possibility out. By
considering the reducible curve $C_1+C_1'$ as a single curve, we
effective replace $S_1$ by a simple ruled surface. Thus we would
reduce Figure~\ref{fig:nd2} to Figure~\ref{fig:one}. That is, the case
$\su(2k+1)$ is reduced to $\su(2k)$ and so we get $\sp(k)$ under
monodromy.

At this point therefore we do not really seem to know what the gauge
algebra is. The geometry seems to suggest $\su(k+1)$ or $\sp(k)$ while
the outer automorphism argument suggests $\sp(k)$ or $\so(2k+1)$.
We will now give various arguments 
in support of 
the following 
assertion.
\begin{mainassertion} \label{conj:x}
For an F-theory compactification on an elliptic threefold with a curve of
$\mathrm{I}_{2k+1}$ fibers (which locally suggests a symmetry of
$\su(2k+1)$) with $\Z_2$ monodromy, the resulting gauge symmetry is
$\sp(k)$ (provided that the R-R fields are set to ``zero'').
\end{mainassertion}
This assertion corrects some statements which appeared in earlier
literature where it had been assumed the resulting gauge symmetry was
$\so(2k+1)$ for reasons we discussed above. As mentioned in the
introduction one can show that the spectrum of various F-theory
models, such as point-like instantons on a $D_n$ singularity, is
anomaly free for $\sp(k)$ but inevitably would have anomalies in some cases
had the gauge algebra contained $\so(2k+1)$.\footnote{See the
footnote in section 4 of \cite{In:po} for a full description; further
calculations of anomaly cancellation conditions in \cite{GM:} also support
our Main Assertion.}

Note that the outer automorphism of $\SU(2k+1)$ which yields $\sp(k)$
actually has order $4$, since its square is conjugation by the matrix
\begin{equation}
 \bar{h} h = h^2 
= \begin{pmatrix} -I&0&\fkm0\\ \fkm0&1&\fkm0\\ \fkm0&0&-I\end{pmatrix},
\end{equation}
where $I$ is the $k\times k$ identity matrix; $h^2$ is not a central
element of $\SU(2k+1)$.  This outer automorphism of course still induces
the required reflection
of the Dynkin diagram, as we explained near the end of section \ref{s:lcy}.

This modifies the analysis which led to eq.~\eqref{eq:rul1} as follows.
We let the outer automorphism of order $4$ act on $\mf{s}$ and decompose
into eigenspaces:
\begin{equation}
  \Ad(\mf{s}) = \Ad(\mf{g})\oplus V_{-} \oplus V_i \oplus V_{-i}  ,
                                                       \label{eq:rul1bis}
\end{equation}
each of which will be a representation of $\mf{g}$ (possibly reducible).
As before, the eigenspaces for eigenvalues $\pm1$ can be accounted for by
certain positive roots which are left invariant under the involution and by
certain pairs of positive roots which are exchanged under that involution.
The moduli space for the former is $M_1$ and for the latter is $M_2$; when
we consider the quantization of the D2-branes wrapped on the corresponding
curves, we find $2g(M_1)$ half-hypermultiplets for each of the invariant
roots, and $2g(M_2)$ half-hypermultiplets for each of the pairs.  Since
each pair contributes to both the $+1$ and $-1$ eigenspace, this adds up to
a total of  $2g(M_1)$ half-hypermultiplets in the adjoint
representation of $\mf{g}$, and $2g(M_2)-2g(M_1)$ half-hypermultiplets in the
representation $V_-$.  

We have yet to account for the representations $V_i$ and $V_{-i}$.  In
fact, these are the roots which contain either $C$ or $C'$, which---as we
have argued above---cannot occur as wrapped D2-branes at the generic point
of the parameter curve $M_1$ if we are to reproduce the Cartan matrix
compatible with
our Main Assertion (or at least, such wrapped branes cannot produce vector
multiplets).\footnote{We remain
mystified as to the exact mechanism 
which obstructs D2-branes from wrapping these unions of curves, or which
removes the vector multiplets from the spectrum of the wrapped branes.
Note that 
Freed and Witten \cite{FW:anomD} have observed obstructions in 
D-branes related to anomalies.}  However, as we will see in
section~\ref{ss:small}, the 
representations $V_i$ and $V_{-i}$ {\it do}\/ occur in the hypermultiplet
spectrum---perhaps because at the branch points of the map $M_2\to M_1$,
$C$ and $C'$ are 
identified and there is no apparent obstruction to wrapping the
D2-brane there.  

To be more concrete concerning the case at hand, with $\mf{s}=\su(2k+1)$,
$\mf{g}=\sp(k)$, and the outer automorphism determined by the $h$ in
eq.~\eqref{eq:h}, we have
\begin{equation}
V_-=\Lambda^2\mathbb{C}^{2k}=(\Lambda^2\mathbb{C}^{2k})_0\oplus \mathbb{C},
\end{equation}
the second exterior power of the fundamental of $\sp(k)$ (which has a
trivial one-dimensional summand), and
\begin{equation}
V_i\cong V_{-i} \cong \mathbb{C}^{2k},
\end{equation}
the fundamental representation of $\sp(k)$.

Thus, the predicted spectrum is: 
\begin{itemize}
\item $g(M_1)$ hypermultiplets in the adjoint representation
\item $g(M_2)-g(M_1)$ hypermultiplets in the second exterior power
representation (including its trivial summand), and
\item additional hypermultiplets in the fundamental representation.
\end{itemize}
In fact, an anomaly calculation \cite{GM:} predicts that there will be
precisely $2g(M_1)-2+\frac32b=2(g(M_2)-g(M_1))+\frac12b$ such
hypermultiplets.  One 
possible interpretation of this formula is that there are two fundamentals
($V_i$ and $V_{-i}$) associated to the parameter curve $M_2$ and an
additional half-fundamental at each branch point.\footnote{There are other
possible interpretations; for example, one can form the degree four cover
$M_3$ of $M_1$ which corresponds to the order four element of $SU(2k+1)$,
and express things in terms of the genus of $M_3$.}

%%%%%%%%%%%%
\subsection{The case of $\su(3)\to\sp(1)$.} \label{ss:small} 

A Kodaira type IV fiber would intrinsically produce an $\su(3)$ gauge
symmetry but monodromy may act on this fiber producing the case of
interest. At first sight this might not look like such a good candidate for
examination since $\su(2)\cong\sp(1)\cong\so(3)$! However, the
hypermultiplet spectrum will allow us to distinguish the cases.

Consider the case of amassing point-like $E_8$-instantons on an
orbifold point of a K3 surface along the lines analyzed in
\cite{AM:po}. We will be interested in the case of four instantons and six
instantons on a $\C^3/\Z_3$ quotient singularity. From result~3 and
figure~7 of \cite{AM:po} we may deduce the spectrum without
encountering any difficulties. In the case of four instantons, the $\Z_3$
singularity may actually be partially resolved to a $\Z_2$
singularity without affecting the particle spectrum. This $\Z_2$
singularity may then be effected by a ``vertical'' line of
$\mathrm{I}_2$ fibers (in the notation of \cite{AM:po}). The six
instanton case is effected by a vertical line of $\mathrm{I}_3$
fibers. The results are
\begin{itemize}
  \item
  For four point-like $E_8$-instantons on a $\Z_3$ singularity we have a
  nonperturbative enhanced gauge algebra of $\su(2)$ with
  hypermultiplets in four $\mathbf{2}$ representations.
  \item
  For six point-like $E_8$-instantons on a $\Z_3$ singularity we have a
  nonperturbative enhanced gauge algebra of
  $\su(2)\oplus\su(3)\oplus\su(2)$ with hypermultiplets as
  $(\mathbf{2},\mathbf{1},\mathbf{1})\oplus(\mathbf{2},\mathbf{3},\mathbf{1})
  \oplus(\mathbf{1},\mathbf{3},\mathbf{1})
  \oplus(\mathbf{1},\mathbf{3},\mathbf{1})
  \oplus(\mathbf{1},\mathbf{3},\mathbf{2})
  \oplus(\mathbf{1},\mathbf{1},\mathbf{2})$.
\end{itemize}

We may also produce exactly the same physics by using a vertical line
of type IV fibers. The configurations of curves of Kodaira fibers in
the base of the elliptic fibration is shown in Figure~\ref{fig:IV} for
the cases of four and six instantons respectively. 
These diagrams are again similar to those presented in \cite{AM:po}
and represent the situation after the base has been blown up the
requisite number of times. The short curved
lines represent fragments of the the curve of $\mathrm{I}_1$ fibers.

\iffigs
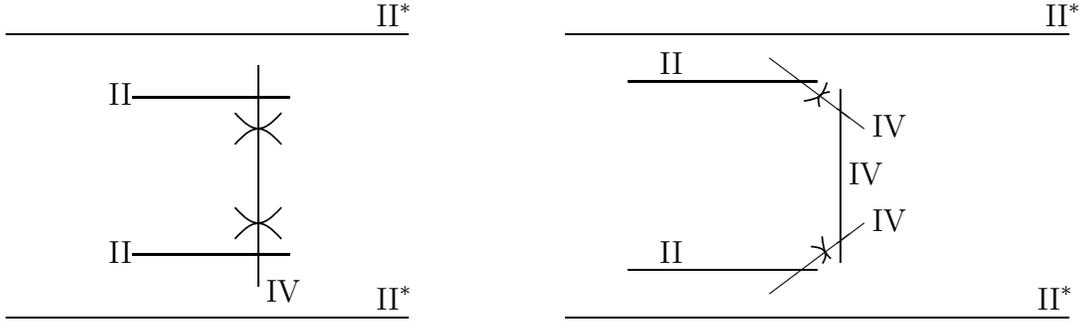
\begin{figure}
\begin{center}
\setlength{\unitlength}{0.00054200in}%
\begin{picture}(10148,2967)(365,-2773)
\thinlines
\put(377,-61){\line( 1, 0){3824}}
\put(1577,-661){\line( 1, 0){1500}}
\put(2777,-361){\line( 0,-1){2100}}
\put(1577,-2161){\line( 1, 0){1500}}
\put(377,-2761){\line( 1, 0){3824}}
\put(5701,-61){\line( 1, 0){4800}}
\put(5701,-2761){\line( 1, 0){4800}}
\put(6301,-511){\line( 1, 0){1800}}
\put(6301,-2311){\line( 1, 0){1800}}
\put(7651,-2536){\line( 4, 3){900}}
\put(8326,-586){\line( 0,-1){1650}}
\put(7651,-286){\line( 4,-3){900}}
\put(2552,-811){
}
\put(2552,-1111){
}
\put(2552,-1711){
}
\put(2552,-2011){
}
\put(7976,-698){
}
\put(8126,-751){
}
\put(8149,-1988){
}
\put(8044,-2086){
}
\put(1352,-2236){\makebox(0,0)[lb]{\smash{$\mathrm{II}$}}}
\put(1352,-736){\makebox(0,0)[lb]{\smash{$\mathrm{II}$}}}
\put(2852,-2611){\makebox(0,0)[lb]{\smash{$\mathrm{IV}$}}}
\put(3901, 14){\makebox(0,0)[lb]{\smash{$\mathrm{II}^*$}}}
\put(3901,-2686){\makebox(0,0)[lb]{\smash{$\mathrm{II}^*$}}}
\put(10201,-2686){\makebox(0,0)[lb]{\smash{$\mathrm{II}^*$}}}
\put(10276, 14){\makebox(0,0)[lb]{\smash{$\mathrm{II}^*$}}}
\put(6601,-436){\makebox(0,0)[lb]{\smash{$\mathrm{II}$}}}
\put(6601,-2236){\makebox(0,0)[lb]{\smash{$\mathrm{II}$}}}
\put(8401,-1486){\makebox(0,0)[lb]{\smash{$\mathrm{IV}$}}}
\put(8626,-1936){\makebox(0,0)[lb]{\smash{$\mathrm{IV}$}}}
\put(8626,-1036){\makebox(0,0)[lb]{\smash{$\mathrm{IV}$}}}
\end{picture}
\end{center}
  \caption{Four and Six Point-like instantons on $\Z_3$ orbifold points.}
  \label{fig:IV}
\end{figure}
\fi

Let us begin with the case of six instantons on the right of 
Figure~\ref{fig:IV}. The lines of type II fibers produce no gauge symmetry
enhancement. 
The upper and lower diagonal lines of type IV fibers
each collide once transversely with a line of type II fibers and once
non-transversely with the curve of $\mathrm{I}_1$ fibers. Actually these
collisions are very similar.\footnote{Indeed for a special choice of moduli,
the line of $\mathrm{I}_1$ fibers can be turned into a line of type II
fibers intersecting the line of type IV fibers transversely.}
Each of these collisions produces $\Z_2$ monodromy in the type IV
fiber producing the geometry of Figure~\ref{fig:nd}. Thus each of
these diagonal lines of type IV fibers produce an $\su(2)$ (or
$\sp(1)$) gauge symmetry.

The remaining vertical line of type IV fibers collides with the other
two lines of type IV fibers. Resolving this collision shows that no
monodromy is induced. Thus this vertical line represents an $\su(3)$
gauge symmetry. An analysis of the collisions shows that there would
be an induced hypermultiplet in the $(\mathbf{3},\mathbf{3})$ of
$\su(3)\oplus\su(3)$ for each collision {\em if there were no
monodromy\/}. Clearly from the desired spectrum above, this
$(\mathbf{3},\mathbf{3})$ must break up as $(\mathbf{2},\mathbf{3})
\oplus(\mathbf{1},\mathbf{3})$ of $\su(2)\oplus\su(3)$. This tells us
immediately that the inclusion $\su(3)\supset\su(2)$ produced by the
action of the monodromy produces a decomposition of the fundamental of
$\su(3)$ via $\mathbf{3}\to\mathbf{2}\oplus\mathbf{1}$. {\em This
rules out the natural embedding $\su(3)\supset\so(3)\cong\su(2)$\/}
for which $\mathbf{3}\to\mathbf{3}$.

We are left with having to account for a hypermultiplet $\mathbf{2}$
in each of the $\su(2)$'s. This must come from the monodromy-inducing
collisions of the diagonal lines of type IV fibers with the lines of
type II and type $\mathrm{I}_1$ fibers. As these collisions are all
the isomorphic locally, each collision must produce a half-hypermultiplet
$\mathbf{2}$. This is in agreement with our comments concerning the 
$V_i$ and $V_{-i}$ representations at the end of the previous subsection.
The collision point is
the point around which the monodromy acts and so it associated with
the location of the curve denoted $D$ in 
Figure~\ref{fig:nd}. 

The choice of associating this $\su(2)$ as the $k=1$ case of $\sp(k)$
or $\su(k+1)$ differs as explained above by whether we view the
positive root of $\su(2)$ as being associated to $C_1$ or to
$C_1+C_1'$. Clearly in the latter case we have $2D=C+C_1$ as divisor
classes and so $D$ naturally generates the $\mathbf{2}$ as
required. If only $C_1$ were identified as the positive root then $D$
would produce nothing new. Therefore we can only correctly identify
the spectrum F-theory in the case of the geometry on the right-hand
side of Figure~\ref{fig:IV} if we take one of roots of the gauge
algebra to be $C_1+C_1'$. That is, there really does appear to be a
rule in string theory which allows $2$-branes to wrap $C_1+C_1'$
together but not $C_1$ or $C_1'$ individually.

We can further verify our picture by considering
the spectrum for four
instantons on the left of Figure~\ref{fig:IV}. 
There are four collisions
with the vertical line of type IV fibers, each producing monodromy.  Thus,
$g(M_2)=1$ and there are $b=4$ branch points.  Following the arguments at
the end of the previous subsection, we thus predict a spectrum
consisting of one hypermultiplet in the
$\Lambda^2\mathbb{C}^2\cong\mathbb{C}$ representation (from $V_-$), and
$2(g(M_2)-g(M_1))+\frac12b=4$
hypermultiplets in the fundamental representation (i.e.\ $V_{\pm i}$).
This precisely agrees with the spectrum found above:  there are four
$\mathbf{2}$'s of $\su(2)$.  Even the $V_-$ representation ``$\mathbf{1}$''
occurs correctly: it
is the deformation {\em \`a la\/} Wilson \cite{Wil:Kc}
of $M_1$, or in physical terms of the heterotic string, it is the
deformation of the $\Z_3$ singularity to a $\Z_2$ singularity which
does not affect the spectrum as noted earlier. 

The rules of $2$-brane wrapping are therefore rather unusual for the
curves $C_1$ and $C_1'$. As observed previously, away from the branch
points, a $2$-brane can {\em never}\/ wrap $C_1$ or $C_1'$ individually.
However, as we have just seen, at 
the branch points where $C_1$ and $C_1'$ coincide, the $2$-brane is allowed
to wrap the curve.  In fact, when this wrapping is taken with both
orientations, a hypermultiplet in the $\mathbf{2}$ of $\su(2)$ is produced
for each branch point.

Since $C_1$ lies in the singular surface $S_1$ it is perhaps not
surprising that the usual rules of $2$-brane wrapping appear to break
down. Anyway, since this same $S_1$ appears as the ``end'' component for
the higher rank gauge groups of this type,  assuming string theory wraps
branes around curves in 
$S_1$ in a similar way in that context, we
arrive at our Main Assertion.

\subsection{Deformation to the $\su(\textrm{even})$ case.}
\label{ss:below}

We can also give a different argument in favor of the $\sp(k)$ gauge group
in case $M_1$ has positive genus.  Let us start with the
case of $\mathrm{I}_{2k+2}$ fibers with $\Z_2$ monodromy.  By 
Theorem~\ref{ieven}, this leads to a $\sp(k+1)$ gauge group and at least
one adjoint hypermultiplet.  We will show in a moment that the 
corresponding $A_{2k+1}$ singularity can be smoothed to an $A_{2k}$ singularity
with $\Z_2$ monodromy.  This corresponds to giving a nonzero vev to
a semisimple element of the adjoint hypermultiplet, and the $\sp(k+1)$
gauge group gets Higgsed to some rank $k$ subgroup $\mf{g}\subset \sp(k+1)$.
We still have to determine what $\mf{g}$ without knowledge of the
$\gu(1)\sp(k+1)$ that acquires a vev.  Clearly $\sp(k)\subset\sp(k+1)$
is possible, so we could have $\mf{g}=\sp(k)$.  We now argue that 
$\mf{g}=\so(2k+1)$ is impossible.

\begin{lemma}
There is no embedding of $\so(2k+1)$ in $\sp(k+1)$ for $k>1$.
\end{lemma}

\medskip\noindent
{\bf Proof}
This argument is due to R.~Zierau.  Suppose that there were an
embedding of $\so(2k+1)$ in $\sp(k+1)$.  Then the fundamental
$2k+2$ dimensional representation $V_{2k+2}$ of $\sp(k+1)$ would restrict
to a representation of $\so(2k+1)$, which necessarily decomposes
as a fundamental representation $V_{2k+1}$ of $\so(2k+1)$ plus a trivial
representation.  The alternating form on $V_{2k+2}$ restricts to a
alternating form on $V_{2k+1}$.  Since $V_{2k+1}$ is odd dimensional,
this form is degenerate.  It's nullspace $W\subset V_{2k+1}$ is invariant
under $\so(2k+1)$, and is a proper subspace since the alternating form
on $V_{2k+1}$ is not identically zero.  This is a contradiction.

\bigskip
The singular surface in question given by eq.~(\ref{eq:prob}) may be
written as
\begin{equation}
  y^2 = x^2z,
\end{equation}
and thought of as a double cover of the $xz$-plane branched along
$z=0$ and doubly along $x=0$. It is the double branching that makes
the surface singular. We may smooth the surface by deforming to
\begin{equation}
  y^2 = x(x-\epsilon)z.
\end{equation}
Now the double branching has been split to $x=0$ and $x=\epsilon$. For
a fixed value of $z$ this process replaces a nodal rational curve by a
smooth rational curve, where the nodal rational curve can be viewed as
two rational curves intersecting transversely at a point. That is,
each pair of intersecting lines in Figure~\ref{fig:nd} is replaced by
a single line and the surface is smoothed.

This smoothing process is remarkably benign at the level of global
geometry. It is often possible to perform it even when the geometry of
the ambient threefold, $X$, is completely smooth at all times. 

We can then derive the $\sp(k)$
gauge symmetry indirectly as follows.  The existence of the deformation
shows that $2$-branes are not allowed to wrap
the individual lines of Figure~\ref{fig:nd}. The deformation converts
each pair of lines into a single new line. Thus if physics is
not discontinuously affected by the deformation, the $2$-branes
contributing to vector particles must only be allowed to 
wrap the pairs of line in Figure~\ref{fig:nd} together.  As we have 
observed in the discussion immediately preceding 
the Main Assertion,
we can now conclude that we do indeed obtain $\sp(k)$.\footnote{Note that
the fact that hypermultiplets may arise from 
wrapping $2$-branes around the individual lines is not compromised by
this argument. When we deform the curve of $A_{2k}$ singularities to a
curve of $A_{2k-1}$ singularities we may affect the geometry of
some points on this curve. Thus hypermultiplets which were ``spread''
over the whole curve of singularities may be localized to isolated
rational curves by this deformation process. Massless vectors cannot
come from such isolated curves.}

There is of course a problem with the proof of 
our Main Assertion
by this argument---there may be global obstructions to such a
deformation. Wilson's criterion suggests that such deformations only occur
when $g(M_1)>0$, where $M_1$ is the base of the fibration of 
Figure~\ref{fig:nd} as a ruled surface.  We address this in part by giving
an example of a deformation when $g(M_1)>0$.

We return to the setup introduced at the end of section \ref{s:hyper}.
Using the notation leading to eq.~(\ref{eq:su2kdef}), we write the equation
\begin{equation}
xy=z^{2k+2}+f_1z^{2k+1}.
\end{equation}
This gives $\su(2k+1)$ with $\Z_2$ monodromy at the generic point of $M_1$.
The deformation is simply
\begin{equation}
xy=z^{2k+2}+f_1z^{2k+1}+\epsilon f_2z^{2k}.
\end{equation}

To make sense of this, we have to say a little more about the blowup.
The $A_{2k}$ blowups are determined by a procedure given in
\cite{KM:weyl} after choosing an ordering of the $2k+1$ factors of
$z^{2k+2}+f_1z^{2k+1}+\epsilon f_2z^{2k} = z^{2k}
(z^2+f_1z+\epsilon f_2)$.  Choosing the $(z^2+f_1z+\epsilon f_2)$ factor to
be in the middle, we obtain the desired geometry.  The last blowup
creates a single ruled surface, which smooths out the singular
component.  It is immediate to see from the description in
\cite{KM:weyl} that for generic $f_i$ this is a smooth deformation
of the desired type.

In this model, we have placed a restriction on the genus and have introduced
localized matter at the zeros of $f_1$.
However, if we are willing to accept that the
process of gauge symmetry enhancement is dictated by local geometry
then this example is enough to justify 
our Main Assertion.

%%%%%%%%%%%%%%%%%%%%%%%%%%%%%%%%%%%%%%%%%%%%%%%%%%%%%%%%%%%%%%%%

\section{Numerical Oddities}   \label{s:num}

Finally we close with a note on the peculiar numerical predictions
dictated by anomaly cancellation in the six-dimensional physics
produced by F-theory compactified on $X$. This has been discussed in
many places before (for example \cite{Sch:a6})  and is often used as a
method of enumerating the 
spectrum of hypermultiplets. Here we have outlined a systematic way of
constructing the hypermultiplet spectrum and so the anomaly constraint
becomes a peculiar numerical property of the geometry of an
elliptic \CY\ threefold.

For completeness we will repeat the anomaly condition here. We
consider an elliptic fibration $\pi:X\to \Sigma$ with a
section. Let $\cG$ be the 
gauge group (or algebra) in six dimensions and $\rho(\Sigma)$ the
Picard number of $\Sigma$. Then anomaly cancellation along the lines of
\cite{AW:grav} yields the following
\begin{equation}
\dim\cG - \sum_i\varepsilon_i\dim R_i = 29\rho(\Sigma) - 302,
                \label{eq:anom}
\end{equation}
where the hypermultiplets fall into representations $R_i$ of $\cG$
and $\varepsilon_i$ is equal to $1$
if the representation is real or $\ff12$ if the representation is
complex or quaternionic (pseudoreal). Note that the trivial
representations also contribute to the sum. These can be determined
from the fact that the number of neutral hypermultiplet moduli are
equal to $h^{2,1}(X)+1$.

As an example consider the extreme case of $\cG\cong E_8^{17}\times 
F_4^{16}\times G_2^{32}\times \SU(2)^{32}$ corresponding to $24$ point-like
$E_8$-instantons on a binary icosahedral quotient singularity in the
heterotic string \cite{AM:po}. The \CY\ threefold for the F-theory
description of this has $\rho(\Sigma)=194$. Applying the methods of
sections \ref{s:hyper} and \ref{ss:small} to this threefold we also
arrive at a spectrum of hypermultiplets of a $(\mathbf{1},\mathbf{2})
\oplus(\mathbf{7},\mathbf{2})$ for each of the $32$ copies of
$G_2\times\SU(2)$.\footnote{The fact that this hypermultiplet spectrum
canceled the anomalies was noted in \cite{CPR:mtor}.} These representations
are quaternionic. Equation  
(\ref{eq:anom}) then reads 
\begin{equation}
  5592 - (\ff12\times 32\times 16 + 12) = 29\times 194-302.
\end{equation}

The anomaly condition in eq.~(\ref{eq:anom}) has been verified in situations
illustrating Theorem \ref{ieven} and our Main Assertion
(see the
footnote in Section 4 of \cite{In:po}).

Note that one may obtain further conditions from the anomaly
cancellation condition. For example one may require the vanishing of
coefficient of each ``$\tr(F^4)$'' term in the anomaly. See for example 
\cite{Sch:a6,Sad:anom}. 

It would be very satisfying to give a purely geometric proof of
eq.~(\ref{eq:anom}) and the other anomaly conditions. (A geometric proof
which covers a wide variety of cases has recently been constructed
\cite{GM:}.) Sadly at present the
origin of this formula without using string 
theory is something of a mystery. Note that the existence of a section
in the fibration $\pi:X\to \Sigma$ is necessary for this to
work. If this requirement is not satisfied then there is no
six-dimensional physics and the condition need not be satisfied \cite{me:lK3}
(for an example, see \cite{BM:Fbun}). 

%%%%%%%%%%%%%%%%%%%%%%%%%%%%%%%%%%%%%%%%%%%%%%%%%%%%%%%%%%%%%%%%

\section*{Acknowledgements}

It is a pleasure to thank A.~Grassi, K.~Intriligator, J.~Morgan, R.~Plesser, 
G.~Rajesh, and R.~Zierau for useful conversations and insights.
P.S.A.\ is supported in part by a research fellowship from the Alfred
P.~Sloan Foundation. 
The work of D.R.M.\ is supported in part by
by NSF grant DMS-9401447.  The work of S.K.\ is supported by NSA grants
MDA904-96-1-0021 and MDA904-98-1-0009, and NSF grant DMS-9311386. S.K.\ also
thanks the Mittag-Leffler Institute for support during the early stages of
this project.

%\bibliographystyle{my-phys}
%\bibliography{string}

\end{document}